\documentstyle[prl,aps,epsfig]{revtex}

\begin{document}
\draft


\title{Doppler cooling of a Coulomb crystal}

\author{Giovanna Morigi$^1$ and J\"urgen Eschner$^2$}

\address{$^1$ Max-Planck-Institut f\"ur Quantenoptik, Hans-Kopfermann-Strasse 1,
D-85748 Garching, Germany\\
$^2$ Institut f\"ur Experimentalphysik, University of Innsbruck,
Technikerstrasse 25/4,
A-6020 Innsbruck, Austria}

\date{\today}

\maketitle

\begin{abstract}
We study theoretically Doppler laser--cooling of a cluster of
2-level atoms confined in a linear ion trap. Using several consecutive steps
of averaging we derive, from the full quantum mechanical master equation, an
equation for the total mechanical energy of the one dimensional
crystal, defined on a coarse-grained energy scale whose grid size is smaller
than the linewidth of the electronic transition. This equation describes the
cooling dynamics for an arbitrary number of ions and in the quantum regime.
We discuss the validity of the ergodic assumption (i.e. that the phase space distribution 
is only a function of energy). From our equation
we derive the semiclassical limit (i.e. when the mechanical motion can be
treated classically) and the Lamb-Dicke limit (i.e. when the size of the
mechanical wave function is much smaller than the laser wavelength). We find
a Fokker-Planck equation for the total mechanical energy of the system, whose
solution is in agreement with previous analytical calculations which were
based on different assumptions and valid only in their specific regimes. Finally, 
in the classical limit we
derive an analytic expression for the average coupling, by light
scattering, between motional states at different energies.
\end{abstract}


\section{Introduction}

The development of laser cooling and trapping techniques in the last decades
has allowed for many spectacular experimental achievements \cite{Nobelex}.
Hand in hand with the experimental work, a theory of laser cooling has been
widely developed, providing a precise description of many experimental
situations \cite{Nobelth,Stenholm}. Yet, the great majority of the
theoretical studies have dealt with the interaction of laser light with
single particles (being the same as many non-interacting particles). The
treatment of a many-body system coupled to light is considerably more
complex, due to the large number of degrees of freedom which results from the
interaction among the constituents of the system.

In this work we investigate laser cooling of a many-body system, taking as a
representative -- and experimentally relevant -- example a one-dimensional
Coulomb cluster, i.e. a crystallized structure of ions which are trapped in a
Paul or Penning trap \cite{Dubin,Cluster1,Cluster2,Cluster3}. We develop a
model for the dynamical behaviour of the crystal's mechanical energy. 
In this system the physical
processes in play are the trapping potential which confines the ions, their
mutual Coulomb interaction, and their interaction with laser light. For
sufficiently small kinetic energy the motion of the ions is properly described
by the collective excitations of the cluster, the so-called harmonic modes of
the crystal, due to the interplay of trapping potential and Coulomb repulsion.
The problem is then characterized by three main frequency scales: (i) the
oscillation frequencies of these modes, which are determined by the trap
frequency $\nu$ and the number of particle $N$ (for ions of equal mass and
charge); (ii) the recoil frequency $\omega_R$ characterizing the exchange of
mechanical energy between radiation and atoms,
$\omega_R=2\pi^2\hbar/m\lambda^2$, with $m$ mass of the ions and $\lambda$
wavelength of the light coupling quasi-resonantly to the electronic
transition, and (iii) the linewidth of the electronic transition $\gamma$,
which characterizes the rate at which photons are scattered by the atoms.
Doppler cooling of particles in a harmonic trap focuses on the regime
$\gamma>\nu$.

The theory is well developed for single trapped particles in specific regimes:
in the semiclassical limit $\gamma\gg\nu,\omega_R$, when the mechanical motion
can be treated classically, and in the Lamb-Dicke limit $\nu\gg\omega_R$, when
the size of the motional wave function is much smaller than the laser
wavelength \cite{Stenholm}. In the semiclassical case, the standard procedure
consists in writing the equations for the system dynamics in the Wigner
representation, adiabatically eliminating the excited electronic state and
then expanding in $\hbar$ up to the second order. In this way a Fokker-Planck
equation in the position and the velocity of the ion is derived
\cite{JavaSten}. In the Lamb-Dicke limit, corresponding to an expansion in
$\omega_R/\nu$ up to the first order, the cooling dynamics are described by a
set of rate equations projected on the electronic ground state and on the
eigenstates of the motion, which can be analytically handled
\cite{Stenholm,Lindberg}. Both treatments yield in the limit $\gamma\gg\nu$
the same final distribution of cooled atoms in energy space, and the same
dependence of the final temperature on the cooling parameters. Our approach
shows how they are special cases of the same equation that we derive for the
mechanical energy of the crystal.

The presence of many mutually interacting particles complicates the treatment
considerably, because the mechanical Hamiltonian, having an increased number
of degrees of freedom, often does not allow for a simple and transparent
solution. An immediately visible effect arises in the spectrum of the
mechanical energy, where the increased number of degrees of freedom is in
general connected with the appearance of quasi-degeneracies and with a dense
distribution of the energy levels \cite{PRAIbk}. The situation is facilitated
again if one restricts the treatment to the Lamb-Dicke regime
\cite{Dipole,Javanainen} but in the more general case we want to discuss,
different steps of simplification can be made as will be shown.

The treatment in this paper focuses on Doppler cooling of an $N$-body
one-dimensional Coulomb crystal, in particular on the type of linear ion
chains obtained in linear Paul traps \cite{Cluster3,Linear}. Here, the
many-body mechanical Hamiltonian can be approximated by a set of harmonic
oscillator modes, all having different frequencies $\nu_1,...,\nu_N$, where
each mode corresponds to a collective excitation of the crystal. Since for
Doppler cooling the rate of photon scattering, and thus the cooling dynamics
is determined by $\gamma$, we define a grid in energy space $\Delta E\ll\hbar
\gamma$ and study the cooling dynamics among the energy shells defined by the grid. 
From this starting
point, with a procedure of successive averaging we derive a rate equation for
the population at each motional energy on the grid. From this result, the
semiclassical limit is recovered by expanding in $\omega_R/\gamma$. In
particular, we get a Fokker-Planck equation for the motional energy describing
cooling of $N$ ions, whose solution agrees with the ones of standard
semiclassical treatments \cite{JavaSten,Javanainen,WineItano,Itano}. 
In the Lamb-Dicke regime,
starting from the full master equation, we discuss the conditions under which
a compact equation for cooling can be found. Finally, we derive an explicit
form of the rate equation in the limit $\gamma\gg \nu$ by evaluating an
analytic expression for the average coupling by light scattering between
motional states at different energies.

Although we restrict the investigation to a one-dimensional Coulomb crystal,
the theory can be directly extended to three dimensions, and the results are
applicable to clusters in both Paul and Penning traps. More in general, the
idea behind our treatment is that -- in incoherent processes -- if the rate
determining the dynamics of interest of the system can be singled out, the
contribution of processes occurring on faster time scales is often well
represented by the average value on the slower time scale, characterizing the
incoherent process.\\
This work is organized as follows. In Section II the model from which our
derivation starts is introduced and discussed. In Section III we discuss and
apply a series of approximation, from which we obtain an equation for the 
cluster's mechanical energy
describing cooling of the crystal. In Section IV we study the equation in the
Lamb-Dicke and Semiclassical limits, and we compare our results with those
obtained in the same limits with other treatments. In Section V we derive an
explicit functional form of the rate equation in the limit $\gamma\gg\nu$.
Finally, we draw our conclusions.

\section{Model}

\noindent In this section we introduce the model which we study throughout the
paper. We first discuss the mechanical properties of a one-dimensional Coulomb
crystal, then the interaction of laser light with the internal electronic transition 
of each ion, and finally the mechanical action of the light on the collective
mechanical degrees of freedom of the crystal.

\subsubsection{Mechanical properties}

\noindent The mechanical potential on an ion cloud in a Paul trap is the sum
of the potential exerted by the trap on each ion and of the Coulomb repulsion
among the ions. Sufficiently far away from the trap electrodes, the
potential of a Paul trap can be considered harmonic and the total potential
for $N$ ions of charge $e$ has the form:
\begin{equation}
\label{Vpot} V=\sum_{j=1}^N
\frac{1}{2}\left(u_{0x}^{j}x_j^2+u_{0y}^{j}y_j^2+u_{0z}^{j}z_j^2\right)
+\frac{1}{2}\sum_{j=1}^N \sum_{k=1,k\neq
j}^N\frac{e^2/4\pi\epsilon_0}{|\vec{r}_j-\vec{r}_k|}.
\end{equation}
where $\vec{r}_j=(x_j,y_j,z_j)$ is the position of ion $j$ and
$u_{0x}^{j},u_{0y}^{j},u_{0z}^{j}$ depend on the trap parameters and on the
mass of the $j$th ion. At sufficiently low temperatures the ions crystallize
around the classical equilibrium positions $\vec{r}_{i}^{(0)}$ which are
solutions of the set of equations $\partial V/\partial
\vec{r}_i|_{\vec{r}_{i}^{(0)}}=0$ \cite{Dubin}. Sufficiently below the
crystallization temperature the motion of the ions around these equilibrium points
is harmonic to good approximation. In this limit, 
the total mechanical
potential can be described by its Taylor series truncated at the second
order in the expansion around $\{\vec{r}_{i}^{(0)}\}$. In a Paul trap with
cylindrical geometry and very steep potential in the radial ($\hat{x}$ and
$\hat{y}$) direction the ions crystallize along the trap ($\hat{z}$-) axis.
In this regime the amplitude of the radial oscillations is much smaller than the
axial ones.  Here, we assume the radial degrees of
freedom to be frozen out such that the motion is one-dimensional along the
$\hat{z}$-axis, and we take the axial potential to be electrostatic, as it is
the case in a linear Paul trap \cite{Linear}. Then the truncated mechanical
potential has the form:
\begin{equation}
V(q_1,...,q_N)=\frac{1}{2}\sum_{j,k=1}^{N} V_{jk} q_j q_k,
\end{equation}
where $V_{jk}$ is a real, symmetric and non-negative matrix (the
explicit form of its elements can be found, for example, in
\cite{James}), while $q_j=z_j-z_j^{(0)}$ are the displacements
from the equilibrium positions $z_j^{(0)}$. Given $m$ mass of 
each ion, the secular equation for the harmonic motion of the crystal
has the form:
\begin{equation}
\label{Secular} \sum_jV_{ij}b_j^{\alpha}=m\nu_{\alpha}^2
b_i^{\alpha}\mbox{~~~with~~~}\alpha=1,...,N
\end{equation}
where $\nu_{\alpha}$ are the eigenvalues and $b_j^{\alpha}$ the
associated eigenvectors, which are complete and orthonormal.\\
In the basis $q_{\alpha}^{\prime}=\sum_i b_i^{\alpha}q_i$ the motion is
described by the Lagrangian for $N$ independent harmonic oscillators of
frequency $\nu_{\alpha}$. Given the canonical momentum
$p_{\alpha}^{\prime}=\dot{q}_{\alpha}^{\prime}$ conjugate to
$q_{\alpha}^{\prime}$, the motion is quantized
by associating a quantum mechanical oscillator with each mode. Denoting by
$a_{\alpha}$ and $a_{\alpha}^{\dagger}$ the annihilation and creation
operators for the mode $\alpha$, respectively, the coordinates $q_i$ are now
written as
\begin{equation}
\label{quant:q} q_i = \sum_{\alpha} \left(b^{-1}\right)_i^{\alpha}
\sqrt{\frac{\hbar}{2m\nu_{\alpha}}} \left(
a_{\alpha}+a_{\alpha}^{\dagger} \right). \label{x_i}
\end{equation}
and the Hamiltonian for the mechanical motion has the form:
\begin{equation}
\label{mec} H_{\rm mec} =\sum_{\alpha}\hbar\nu_{\alpha}
\left(a_{\alpha}^{\dagger}a_{\alpha}+\frac{1}{2}\right)
\end{equation}
The energy eigenstates of each mode $\alpha$ are the number states
$|n_{\alpha}\rangle$ with eigenvalues (energies) ${\mathcal
E}_{n_{\alpha}}=\hbar (n_{\alpha}+\frac{1}{2})\nu_{\alpha}$. In the following
we use states $|{\bf n}\rangle=|n_1,n_2,...n_N\rangle$ to describe the
eigenstates of the motion of the crystal corresponding to the eigenvalues
${\mathcal E}_{\bf n}=\sum_{\alpha}{\mathcal E}_{n_{\alpha}}$.

In summary, $N$ trapped ions crystallized in one dimension can be described by
$N$ harmonic oscillators with frequencies $\nu_1,...,\nu_N$. These frequencies
are solutions of (\ref{Secular}) and, what is important, they are
incommensurate. Therefore the spectrum of mechanical energies of the system
does not exhibit the discreteness and equispacing property of the single
harmonic oscillator spectrum, but shows a dense distribution of levels, and at
sufficiently high energies and large $N$ it assumes a quasi--continuum
character. In this limit we can define the density of states of the system,
$g(E)$, which is a smooth function of the energy and is defined on a grid of
energies $\Delta E$, such that the number of states $D(E)$ contained in the
interval ${\cal F}(E)=[E-\Delta E/2,E+\Delta E/2]$ is $D(E)\gg 1$, and
$D(E)\simeq g(E)\Delta E$. For $N$ modes (ions) $g(E)$ can be evaluated by
solving the following integral:
\begin{eqnarray}
g(E) &=&\frac{1}{(\hbar\nu_1)(\hbar\nu_2)...(\hbar\nu_N)}\frac{\rm d}{{\rm
d}E} \int_0^E {\rm d} E_1
\int_0^{E_1} {\rm d} E_2...\int_0^{E_{N-1}} {\rm d} E_N\nonumber\\
&=&\frac{E^{N-1}}{(N-1)!(\hbar\nu_1)(\hbar\nu_2)...(\hbar\nu_N)}
\label{smoothDensity}
\end{eqnarray}
i.e. by taking the derivative of the ratio between the volume in phase space
of energy $\leq E$ and the volume occupied by a single state,
$(\hbar\nu_1)(\hbar\nu_2)...(\hbar\nu_N)$. In Fig.~1 we plot the number of
states as a function of the total mechanical energy for a chain of three ions,
taking a grid $\Delta E=\hbar\nu_1/5$, and compare the exact value with the
smoothed function $g(E)\Delta E$, with $g(E)$ given by (\ref{smoothDensity}).
\begin{figure}
\begin{center}
\epsfxsize=0.4\textwidth
\epsffile{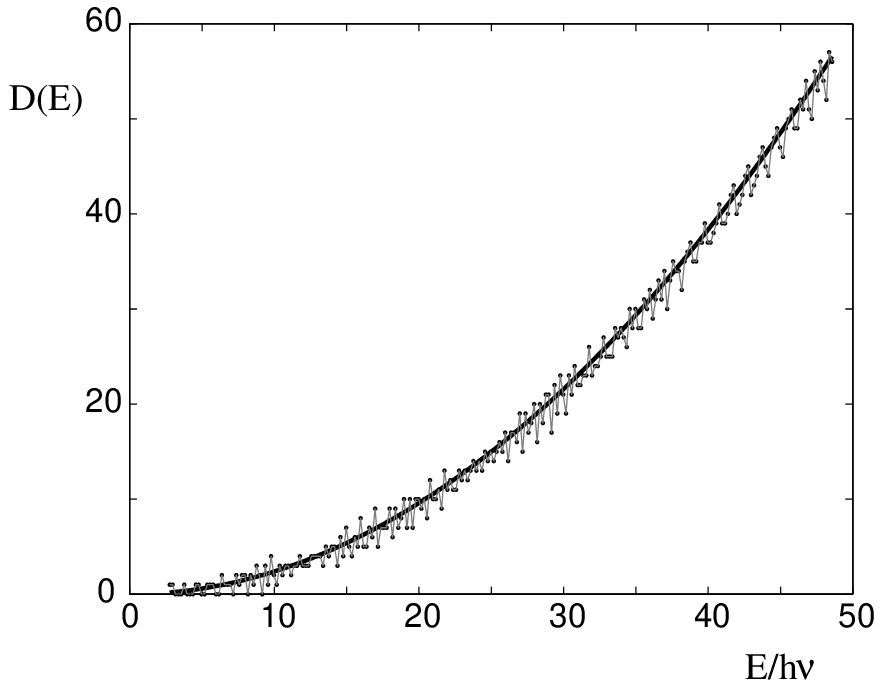} \caption {Number of states $D(E)$ as a function of the
total energy $E/\hbar\nu$ evaluated for three ions on a grid $\Delta
E=\hbar\nu/5$ (dots joined by the grey line); smoothed function $g(E)\Delta E$
of Eq.~(\ref{smoothDensity}) with $N=3$ (black line). The frequencies of the
modes are $\nu_1=\nu$, $\nu_2=1.7321\nu$, $\nu_3=2.4083\nu$. }
\end{center}
\end{figure}

\subsubsection{Interaction with light}

\noindent We consider that laser light with frequency $\omega_L$ and wave
vector $\vec{k}=(k_x,k_y,k_z)$ drives some ions of the crystal, coupling to
their internal two-level transition with electronic ground state
$|\text{g}\rangle$, excited state $|\text{e}\rangle$ and resonance frequency
$\omega_0$. In the Rotating Wave Approximation and in the frame rotating with
the laser frequency, the total Hamiltonian has the form:
\begin{equation}
H = H_{\rm at}  + H_{\rm mec} + H_{\rm AL}.  \label{Hamiltonian}
\end{equation}
Here $H_{\rm at}$ is the Hamiltionian for the internal degrees of
freedom, defined as
\begin{equation}
H_{\rm at}=-\hbar\delta{\sum_{j=1}^{N}}|\mbox{e}\rangle_j\langle
\mbox{e}|,
\end{equation}
where $\delta=\omega_L-\omega_0$ is the detuning of the laser from
the atomic resonance and $j$ labels the ions. $H_{\rm mec}$ is the
mechanical Hamiltonian defined in (\ref{mec}), and $H_{\rm AL}$
describes the interaction between laser and atoms,
\begin{equation}
H_{\rm AL}=\hbar\sum_{\{j\}}\frac{\Omega(z_j)}{2}
[\sigma_j^{+}\text{e}^{-i k_z z_j} + \sigma_j^{-}\text{e}^{i k_z
z_j}], \label{Hal}
\end{equation}
where $\Omega(z_j)$ is the Rabi frequency at the position $z_j$,
$\sigma_j^{+}$, $\sigma_j^{-}$ are the raising and lowering dipole operators,
respectively, for the internal state of the $j^{\rm th}$ ion, while $\{j\}$ is
the set of driven ions. Assuming that the light intensity does not vary
rapidly in the vicinity of $z_j^{(0)}$, we can approximate $\Omega(z_j)\approx
\Omega(z_j^{(0)})=\Omega_j$. The operator $\text{e}^{-ik_zz_j}$ in (\ref{Hal})
represents the mechanical effect of the light interaction, i.e. a shift in
momentum space by one photon recoil which goes along with the excitation.
Using (\ref{quant:q}), its explicit form is:
\begin{equation}
\label{kick} \exp(-i k_z z_j)=\exp\left(-i k_z z_j^{(0)}-
i\sum_{\alpha=1}^N  \eta_j^{\alpha}
(a^{\dagger}_{\alpha}+a_{\alpha}) \right),
\end{equation}
where $\eta_j^{\alpha}$ is the Lamb-Dicke parameter for mode $\alpha$ and ion
$j$, defined as \cite{EPJD}:
\begin{equation}
\label{LambDicke} \eta_{j}^{\alpha} = k_z
\left(b^{-1}\right)_j^{\alpha} \sqrt{\frac{\hbar}{2m\nu_{\alpha}}}.
\end{equation}
For some ions we can have $\Omega_j=0$, which means that not all ions of the
chain are driven. Such condition can be achieved either when the ions are
sufficiently spaced to allow for their individual addressing
\cite{AddressIbk}, or by introducing a different type of ion into the crystal,
such as an isotope of the trapped element \cite{Aahrus}, or a different
species \cite{EPJD,NIST2} whose transition frequency is not resonant with the
laser. In this case, some differences in the mechanical behaviour
arise, which, however, are not relevant for the results that we derive below.

\subsubsection{Master Equation}

\noindent We describe the dynamics of the driven crystal through the master
equation for the density matrix $\rho$ of the $N$--ion system:
\begin{equation}
\frac{\rm d}{{\rm d}t}\rho=-\frac{i}{\hbar}[H,\rho]+L(\rho)
\label{Master0},
\end{equation}
where $H$ has been defined in (\ref{Hamiltonian}), and $L$ is the Liouvillian
describing the incoherent evolution of the system due to its coupling to the
modes of the electromagnetic field:
\begin{equation}
L(\rho)=\frac{\gamma}{2} \sum_{j=1}^N
[2\sigma_j^-\tilde{\rho}_j\sigma_j^+ -\sigma_j^+
\sigma_j^-\rho-\rho\sigma_j^+\sigma_j^-].
\end{equation}
Here $\gamma$ is the decay rate of the internal excited states
$|\text{e}\rangle_j$, and $\tilde{\rho}_j$ is the density matrix
describing the feeding for the ions $j$, shifted in momentum space by
the recoil of the spontaneously emitted photon,
\begin{equation}
\tilde{\rho}_j=\int_{-1}^1 \text{d}(\cos\theta) {\mathcal
N}(\cos\theta) \text{e}^{ik \cos\theta z_j}\rho\text{e}^{-ik
\cos\theta z_j},
\end{equation}
with ${\mathcal N}(\cos\theta)$ being the dipole pattern of the
spontaneous decay, and $k=|\vec{k}|$.

\subsubsection{Low saturation limit}

\noindent In the limit of low saturation, $\Omega\ll\gamma$, the excited
states $|\text{e}\rangle_j$ can be eliminated from (\ref{Master0}) in second
order perturbation theory \cite{Gardiner}. Thereby we obtain a closed
equation for the internal ground state $|{\rm
g}\rangle=\prod_{j=1}^N|\text{g}\rangle_j$, which we project on the basis of
mechanical states $|{\bf n}\rangle$. The equations describing the evolution
of this system are \cite{PRAIbk}:
\begin{eqnarray}
\label{no_neglect} \frac{\rm d}{{\rm d}t} \langle {\bf n}|\rho |{\bf m}\rangle
&=    &-\frac{i}{\hbar}({\mathcal E}_{\bf n}-{\mathcal E}_{\bf m})
\langle {\bf n}|\rho |{\bf m}\rangle \\
&+    & i\sum_{\{j\}}\frac{\Omega_j^2}{4}\sum_{{\bf k},{\bf l}}
\left[\frac{\langle {\bf n}|\text{e}^{ik_zz_j}|{\bf k}\rangle \langle {\bf
k}|\text{e}^{-ik_zz_j}|{\bf l}\rangle} {({\mathcal E}_{\bf k}-{\mathcal E}_{\bf
l})/\hbar-\delta -i\gamma/2} \langle {\bf l}|\rho |{\bf m}\rangle
-\frac{\langle {\bf l}|\text{e}^{ik_zz_j}|{\bf k}\rangle \langle {\bf
k}|\text{e}^{-ik_zz_j}|{\bf m}\rangle} {({\mathcal E}_{\bf k}-{\mathcal E}_{\bf
l})/\hbar-\delta +i\gamma/2} \langle {\bf n}|\rho |{\bf l}\rangle
\right]\nonumber\\
&+    &\gamma\sum_{\{j\}}\frac{\Omega_j^2}{4} \sum_{{\bf k},{\bf j},{\bf
r},{\bf s}}\int_{-1}^{1}\text{d}(\cos\theta){\mathcal
N}(\cos\theta) \langle {\bf n}|\text{e}^{ik\cos\theta z_j}|{\bf
k}\rangle \langle {\bf k}|\text{e}^{-ik_zz_j}|{\bf r}\rangle
\langle {\bf s}|\text{e}^{ik_zz_j}|{\bf j}\rangle \langle {\bf
j}|\text{e}^{-ik\cos\theta z_j}|{\bf m}\rangle
\langle {\bf r}|\rho |{\bf s}\rangle \nonumber\\
&\cdot& \left[\frac{1}{[({\mathcal E}_{\bf j}-{\mathcal E}_{\bf s}-{\mathcal
E}_{\bf k}+{\mathcal E}_{\bf r})/\hbar +i\gamma] [({\mathcal E}_{\bf
j}-{\mathcal E}_{\bf s})/\hbar -\delta -i\gamma/2]} + \frac{1}{[({\mathcal
E}_{\bf k}-{\mathcal E}_{\bf r}-{\mathcal E}_{\bf j}+{\mathcal E}_{\bf
s})/\hbar-i\gamma] [({\mathcal E}_{\bf k}-{\mathcal E}_{\bf r})/\hbar-\delta
+i\gamma/2]} \right]\nonumber.
\end{eqnarray}
Here, the coefficients describing the coupling among 
the populations, $\langle {\bf n}|\rho |{\bf n}
\rangle$, and the coherences, $\langle {\bf n}|\rho |{\bf m}
\rangle$, are proportional to the Franck--Condon coefficients,
which have the form:
\begin{eqnarray}
\langle  {\bf l}|\text{e}^{-ikz_j}| {\bf n}\rangle
&=&\prod_{\alpha=1}^N\langle  l_{\alpha}|
\text{e}^{-i\eta_j^{\alpha}(a_{\alpha}+a_{\alpha}^{\dagger})}|
n_{\alpha}\rangle\nonumber\\
&=&\prod_{\alpha=1}^N\exp\left(-\frac{{\eta_j^{\alpha}}^2}{2}\right)
\sqrt{\frac{r_{\alpha}!}{(r_{\alpha}+|l_{\alpha}-n_{\alpha}|)!}}
(i\eta_j^{\alpha})^{|l_{\alpha}-n_{\alpha}|} {\rm
L}_{r_{\alpha}}^{|l_{\alpha}-n_{\alpha}|} \left({\eta_j^{\alpha}}^2\right)~,
\label{fc}
\end{eqnarray}
where $r_{\alpha}=\text{min}(l_{\alpha},n_{\alpha})$, and ${\rm
L}_{r_{\alpha}}^{|l_{\alpha}-n_{\alpha}|}\left(x\right)$ is a generalized
Laguerre polynomial. They represent the probability amplitude of a transition from 
the initial motional state $|{\bf n}\rangle$ to the final motional state 
$|{\bf l}\rangle$ by absorption or emission of a photon.

\section{An ergodic equation for laser cooling of the crystal}

\noindent In the following, starting from the full quantum mechanical
Eqs.~(\ref{no_neglect}) and using consecutive steps of averaging, we derive
an equation for Doppler cooling of an ion crystal. This equation will be
ergodic, in the sense that it describes the dynamics of the system in terms
of the population $P(E)$ at the crystal's mechanical energy $E$.

In the Doppler cooling limit (which implies $\gamma>\nu_{\alpha}$ for all
$\alpha=1,...,N$, and $\delta=\mbox{O}(\gamma)$) we define a grid of energies
$\Delta E$ such that in an interval ${\cal F}(E)=[E-\Delta E/2,E+\Delta E/2]$
the number of states $D(E)\gg 1$, and the Lorentzian describing the resonant
response of the atom varies infinitesimally on each interval. The resulting
shells of energies can be visualized as shown in Fig.~2 for the case of a
two-ion crystal, where each axis represents the energy of one mode, and each
point is a state $|{\bf n}\rangle = |n_1,n_2\rangle$. Here, the
states with total energy ${\mathcal E}$ fall into the shell of width $\Delta
E$ along the line $E=\hbar \nu_1(n_1+\frac{1}{2})+\hbar
\nu_2(n_2+\frac{1}{2})$.
\begin{figure}
\begin{center}
\epsfxsize=0.3\textwidth
\epsffile{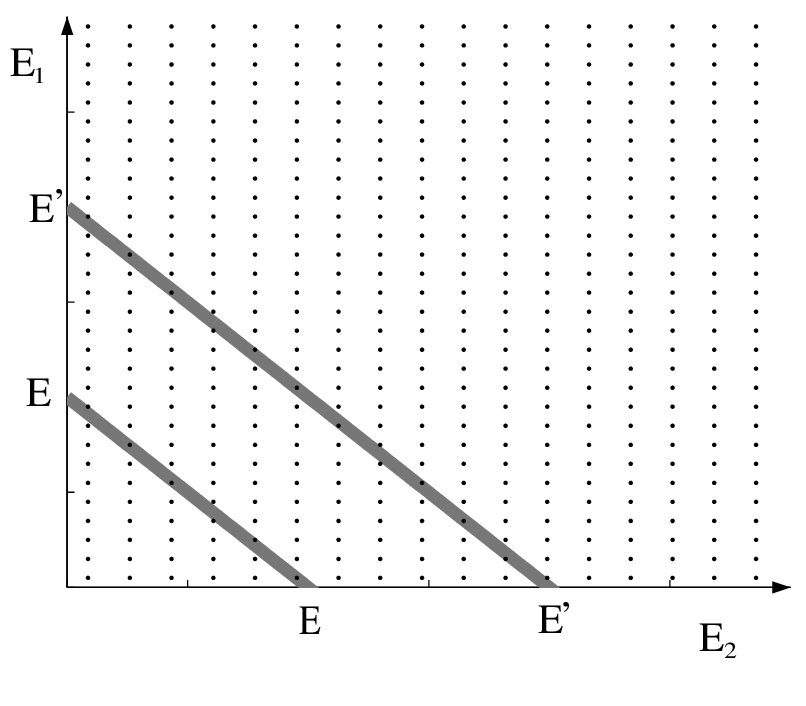} \caption{ Coarse grained energy space for the case of two
modes of frequency $\nu_1$, $\nu_2$. The points are the states with energy
${\mathcal E}_{\bf n}={\mathcal E}_{n_1}+{\mathcal E}_{n_2}$. The broad lines
represent two energy shells ${\cal F}(E)$ and ${\cal F}(E^{\prime})$. }
\end{center}
\end{figure}
We can now decompose the sums over the motional states appearing in
(\ref{no_neglect}) as follows:
\begin{equation}
\sum_{\bf k} = \sum_{E_{\bf k}}{\sum_{{\bf k}}}^{\prime} \mbox{~~where~~}
{\sum_{{\bf k}}}^{\prime} \equiv \sum_{{\bf k}~:~ {\mathcal E}_{\bf k} \in
{\cal F}(E_{\bf k})} \nonumber
\end{equation}
where now $E_{\bf k}$ is one energy value on the grid, and ${\cal F}(E_{\bf
k})=[E_{\bf k}-\Delta E/2,E_{\bf k}+\Delta E/2]$ is the energy interval
centered in $E_{\bf k}$ with width $\Delta E$. Using these definitions we
rewrite the equations for the populations in (\ref{no_neglect}) and sum over
all states that lie within the same energy shell ${\cal F}(E_{\bf k})$:
\begin{eqnarray}
&&\frac{\rm d}{{\rm d}t} {\sum_{{\bf n}}}^{\prime}\langle {\bf
n}|\rho |{\bf n}\rangle =\nonumber \\
&&-\gamma
\sum_{\{j\}}\frac{\Omega_j^{2}}{4}\sum_{E_{\bf k}}\frac{1}{\left[
(E_{\bf k}-E_{\bf n})/\hbar-\delta \right]
  ^{2}+\gamma^2/4 }
{\sum_{{\bf n},{\bf k}}}^{\prime} |\langle {\bf
n}|\text{e}^{ik_zz_j}|{\bf k}\rangle |^{2}
\langle {\bf n}|\rho |{\bf n}\rangle \label{line1}\\
&& -          \sum_{\{j\}}\frac{\Omega_j^2}{4} \sum_{E_{\bf
k},E_{\bf l}}\Im\left\{\frac{1}{(E_{\bf k}-E_{\bf l})/\hbar-\delta
-i\gamma/2} {\sum_{{\bf n},{\bf k}}}^{\prime}{\sum_{{\bf l},{\bf
l}\neq {\bf n}}}^{\prime} \langle {\bf n}|\text{e}^{ik_zz_j}|{\bf
k}\rangle \langle {\bf k}|\text{e}^{-ik_zz_j}|{\bf l}\rangle
\langle {\bf l}|\rho |{\bf n}\rangle\right\}\label{line2}\\
&& +          \gamma\sum_{\{j\}}\frac{\Omega_j^2}{4}
\int_{-1}^{1}\text{d}(\cos\theta){\mathcal
N}(\cos\theta)\sum_{E_{\bf k},E_{\bf r}}\Bigg[ \frac{1}{\left[(E_{\bf
k}-E_{\bf r})/\hbar-\delta \right] ^{2}+\gamma ^{2}/4} {\sum_{{\bf
n},{\bf k},{\bf r}}}^{\prime} |\langle {\bf
n}|\text{e}^{ik\cos\theta z_j}|{\bf k}\rangle |^{2}|\langle {\bf
k}| \text{e}^{-ik_zz_j}|{\bf r}\rangle |^{2}\langle {\bf r}|\rho
|{\bf r}\rangle
\label{line3}\\
&& +          \sum_{E_{\bf j},E_{\bf s}} {\sum_{{\bf n},{\bf
k},{\bf s}}}^{\prime} {\sum_{{\bf r},{\bf j} \atop {\bf j}\neq {\bf
k}~{\rm or}~{\bf r}\neq {\bf s}}}^{\prime} \Re\left\{ \frac{\langle
{\bf n}|\text{e}^{ik\cos\theta z_j}|{\bf k}\rangle\langle {\bf k}|
\text{e}^{-ik_zz_j}|{\bf r}\rangle \langle {\bf
s}|\text{e}^{ik_zz_j}|{\bf j}\rangle\langle {\bf j}|
\text{e}^{-ik\cos\theta z_j}|{\bf n}\rangle } {[(E_{\bf j}-E_{\bf s}
-E_{\bf k}+E_{\bf
r})/\hbar+i\gamma] [(E_{\bf j}-E_{\bf s})/\hbar-\delta
-i\gamma/2]}\langle {\bf r}|\rho
|{\bf s}\rangle\right\}\Bigg]\label{line4},
\end{eqnarray}
where we have separated the terms involving the populations, (\ref{line1}) and
(\ref{line3}), from the ones which involve the coherences, (\ref{line2}) and
(\ref{line4}). We now compare the term (\ref{line1}), the loss rate of the
population, with the coupling to the coherences in (\ref{line2}): Line
(\ref{line1}) is proportional to the sum of moduli square of
Franck-Condon-coefficients, thus it is the sum of positive terms. The sum in
line (\ref{line2}) adds up coefficients with alternating signs, as can be
verified from Eq.~(\ref{fc}). For $D(E)\gg 1$ we can therefore safely assume
that
\begin{equation}
\sum_{\{j\}}{\sum_{{\bf n}}}^{\prime}{\sum_{{\bf k}}}^{\prime} |\langle
{\bf n}|\text{e}^{ik_zz_j}|{\bf k}\rangle |^{2} \langle {\bf
n}|\rho |{\bf n}\rangle\gg |\sum_{\{j\}}{\sum_{{\bf n},{\bf k},{\bf
l}\neq {\bf n}}}^{\prime} \langle {\bf n}|\text{e}^{ik_zz_j}|{\bf
k}\rangle \langle {\bf k}|\text{e}^{-ik_zz_j}|{\bf l}\rangle
\langle {\bf l}|\rho |{\bf n}\rangle| \label{RandomPhase}
\end{equation}
This is equivalent to a Random Phase Approximation. On the basis of this
consideration the coupling between populations and coherences in (\ref{line2})
may be neglected in comparison to the loss rate described by the term in
(\ref{line1}). Analogously we can neglect the terms in (\ref{line4}) in
comparison with (\ref{line3}), and we obtain the following set of rate
equations:
\begin{eqnarray}
\label{rate2}
\frac{\rm d}{{\rm d}t} {\sum_{\bf n}}^{\prime} \langle {\bf
n}|\rho |{\bf n}\rangle
& & = -\gamma \frac{\Omega^{2}}{4}\sum_{\{j\}}\sum_{E_{\bf k}}
\frac{1}{\left[ (E_{\bf k}-E_{\bf n})/\hbar - \delta \right]^{2}+\gamma
^{2}/4}  {\sum_{\bf n}}^{\prime} \langle {\bf n}|\rho
|{\bf n}\rangle {\sum_{\bf k}}^{\prime} |\langle {\bf
n}|\text{e}^{ik_zz_j}|{\bf
k}\rangle |^{2}  \\
 + \gamma \frac{\Omega^{2}}{4}\sum_{\{j\}}
& &
\int_{-1}^{1}\text{d}(\cos\theta){\mathcal
N}(\cos\theta)\sum_{E_{\bf k},E_{\bf_r}} \frac{1}{\left[ (E_{\bf
k}-E_{\bf r})/\hbar-\delta \right] ^{2}+\gamma ^{2}/4}
{\sum_{\bf r}}^{\prime} \langle {\bf r}|\rho |{\bf r}\rangle
{\sum_{\bf k}}^{\prime} |\langle {\bf k}|\text{e}^{-ik_zz_j}|{\bf
r}\rangle |^{2} {\sum_{\bf n}}^{\prime} |\langle {\bf
n}|\text{e}^{ik\cos\theta z_j}|{\bf k}\rangle |^{2}\nonumber,
\end{eqnarray}
where all Rabi frequencies have been set equal, $\Omega_j=\Omega$. This
simplifies the discussion in the next subsection, but the more general case of
different $\Omega_j$ can be treated with the same methods under conditions
which will be discussed below.

\subsection{Ergodic hypothesis}

\noindent Equation~(\ref{rate2}) describes how laser light couples to the internal
electronic transition and exchanges mechanical energy with the crystal. The
probability amplitude for the radiative event to occur is weighted by a
Lorentzian distribution, which below saturation has width $\gamma$, and which
is a function of the difference between the mechanical energies of the crystal
before and after the scattering. The mechanical effect of light on the motion of the crystal 
is described by the operator (\ref{kick}), and the probability for the crystal
initially in state $|{\bf n}\rangle$ to be scattered into state $|{\bf
k}\rangle$ by the absorption or emission of a laser photon is given by the
modulus square of the Franck-Condon coefficient, Eq.~(\ref{fc}). When we
regard the modulus square of Eq.~(\ref{fc}) as a distribution over the states
$|{\bf k}\rangle$ after an absorption or emission event, given the initial
state $|{\bf n}\rangle$, then the average motional energy transferred to the
ion cluster and its variance are calculated as (see Appendix A):
\begin{eqnarray}
&\langle& {\mathcal E}_{\bf k}-{\mathcal E}_{\bf n}\rangle_{\bf k} = \sum_{\bf
k}({\mathcal E}_{\bf k}-{\mathcal E}_{\bf n})~ |\langle {\bf n}|
\mbox{e}^{ik\cos\theta z_j}|{\bf k}\rangle |^{2} = \hbar\omega_R \cos^2\theta,
\label{Center1}\\
&\langle& ({\mathcal E}_{\bf k}-{\mathcal E}_{\bf n})^2 \rangle_{\bf k} =
\hbar \omega_R \cos^2\theta \frac{2 {\mathcal E}_{\bf n}}{N} + (\hbar\omega_R
\cos^2\theta)^2, \label{width}
\end{eqnarray}
where $\omega_R=\hbar k^2/2m$ is the recoil energy. Thus, the first and
second moments of this distribution depend on the (single-ion) recoil energy
$\omega_R$ and on the energy per ion of the initial state, but not on the
details of the quantum state $|{\bf n}\rangle$.

Guided by this result, we will now make the approximation that in the limit
of Doppler cooling, $\gamma > \nu_j$, and for sufficiently large density of
states, $D(E) \gg 1$, the oscillations of the Franck-Condon coefficients with
the vibrational numbers play a negligible role, whereas the average
properties determine the cooling dynamics. We formulate this assumption by
first introducing an average Franck-Condon coupling between shells of energy
$E_{\bf n}$ and $E_{\bf k}$,
\begin{equation}
\label{averageFC} Q^{(k)}(E_{\bf n},E_{\bf k}) = \frac{1}{N
D(E_{\bf n}) D(E_{\bf k})} \sum_{j=1}^N {\sum_{\bf n}}^{\prime}
{\sum_{\bf k}}^{\prime} |\langle {\bf k}| \text{e}^{-ikz_j}|{\bf
n}\rangle |^{2},
\end{equation}
and then writing
\begin{equation}
\label{ergo:1} \frac{1}{N} \sum_{j=1}^N {\sum_{\bf k}}^{\prime}
|\langle {\bf k}| \text{e}^{-ikz_j}|{\bf n}\rangle |^{2} \approx
D(E_{\bf k}) Q^{(k)}(E_{\bf n},E_{\bf k}),
\end{equation}
thereby neglecting the dependence of the LHS on the details of the state
$|{\bf n}\rangle$. We will discuss this assumption and possible alternatives
in the next section. It will be one of the important results later to
determine an explicit expression for $Q^{(k)}(E_{\bf n},E_{\bf k})$ in
Eq.~(\ref{averageFC}). Furthermore, in (\ref{ergo:1}) all ions of the crystal
are assumed to be driven. In the next section we will discuss the case in which
only a set is driven.

Applying (\ref{ergo:1}) to Eq.~(\ref{rate2}), in the limit in
which all ions are driven, we obtain
\begin{eqnarray}
\frac{\rm d}{{\rm d}t} {\sum_{\bf n}}^{\prime} \langle {\bf
n}|\rho |{\bf n}\rangle &=& -N \gamma \frac{\Omega^{2}}{4}\sum_{E_{\bf k}}
\frac {D(E_{\bf k}) Q^{(k_z)}(E_{\bf n},E_{\bf k})} {\left[ (E_{\bf
k}-E_{\bf n})/\hbar - \delta \right]^{2}+\gamma ^{2}/4} {\sum_{\bf
n}}^{\prime} \langle {\bf n}|\rho |{\bf n}\rangle
\nonumber \\
&+& N \gamma \frac{\Omega^{2}}{4} \int_{-1}^{1}\text{d}(\cos\theta)
{\mathcal N}(\cos\theta) \sum_{E_{\bf k},E_{\bf_r}} \frac
{D(E_{\bf k})Q^{(k_z)}(E_{\bf k},E_{\bf r}) D(E_{\bf n})
Q^{(k\cos\theta)}(E_{\bf n},E_{\bf k})} {\left[ (E_{\bf k}-E_{\bf
r})/\hbar-\delta \right] ^{2}+\gamma ^{2}/4} {\sum_{\bf r}}^{\prime}
\langle {\bf r}|\rho |{\bf r}\rangle. \label{rate3}
\end{eqnarray}
In order to obtain the second line of (\ref{rate3}) from the second line of
(\ref{rate2}) we have made a further approximation: First we write
\begin{equation}
\sum_j |\langle {\bf k}|\text{e}^{-ik_zz_j}|{\bf r}\rangle |^{2}
|\langle {\bf n}|\text{e}^{ik\cos\theta z_j}|{\bf k}\rangle |^{2}
= \sum_j |\langle {\bf k}|\text{e}^{-ik_zz_j}|{\bf r}\rangle |^{2}
\sum_l|\langle {\bf n}|\text{e}^{ik\cos\theta z_l}|{\bf k}\rangle
|^{2}~\delta_{jl},
\end{equation}
and then we set $\delta_{jl} \to 1/N$. This corresponds to the assumption
that, in the regime we are considering, the mechanical effect of
the process of absorption + emission of a photon by one ion is equivalent to
absorption by one ion and emission by another ion, weighted by the
probability for the two ions to be the same.

By defining the population densities $P(E,t)$ of the energy shells through
\begin{equation}
\label{P:E} \Delta E P(E_{\bf n},t) = {\sum_{{\bf n}}}^{\prime} \langle {\bf
n }|\rho| {\bf n} \rangle~,
\end{equation}
such that $\int_0^{\infty}{\rm d}E P(E,t)=1$, and by using $D(E) = g(E)\Delta
E$, we finally arrive at a rate equation as a function of the motional
energy:
\begin{eqnarray}
\label{rateEnergy}
\frac {\rm d}{{\rm d}t} P(E,t) &=& -\gamma \frac{\Omega^{2}}{4} N
\int_0^{\infty}\text{d} E_1 \frac {g(E_1) Q^{(k_z)}(E,E_1)}
{\left[ (E_1-E)/\hbar -\delta \right] ^{2}+\gamma ^{2}/4} P(E,t)
 \\
&+& \gamma \frac{\Omega^{2}}{4}N \int_{-1}^{1} \text{d}(\cos\theta)
{\mathcal N}(\cos\theta) \int_0^{\infty} \text{d} E_1
\int_0^{\infty} \text{d} E_2 \frac {g(E) Q^{(k\cos\theta)}(E_1,E)
g(E_1) Q^{(k_z)}(E_2,E_1)} {\left[ (E_1-E_2)/\hbar-\delta \right]
^{2}+\gamma ^{2}/4} P(E_2,t) \nonumber,
\end{eqnarray}
where we have replaced the sums over the energy by integrals, valid when the
average energy $\langle E(t)\rangle \gg \Delta E$. The two parts of the rate
equation (\ref{rateEnergy}) show how, after the various steps of averaging,
the total population of a shell with energy $E$ changes in time: Population
is lost by excitation to shells with energy $E_1$ and subsequent emission
into any shell, and population originally at $E_2$ is excited to a shell at
$E_1$ and then scattered into a shell at energy $E$.\\

It should be noted that the restriction to a one-dimensional crystal enters
into the number of modes and into the geometry of the laser beams, as well as
into the pattern of the emitted radiation. However, the application of the
``Random Phase Approximation'' (\ref{RandomPhase}) and of assumption (\ref{ergo:1}), 
which are at the
basis of our derivation, are in no way restricted by the dimensionality of the
problem. Actually, an extension to three dimensions would endorse the two
approximations, since the number of modes increases, and with it the density
of states in the spectrum of the motional energies. Therefore, an equation for
the total motional energy of the crystal of the form of (\ref{rateEnergy}) can
be derived in three dimensions using the same considerations done here for the
one-dimensional case.

\subsection{Discussion}

\subsubsection{Ergodic assumption}

\noindent Since the distribution of the total mechanical energy over the
system is sufficient for describing the cooling process, assumption
(\ref{ergo:1}), leading to Eq.~(\ref{rateEnergy}), simplifies the description
of the laser cooling dynamics of the crystal significantly by reducing
dramatically the dimensionality of the problem. Equation~(\ref{rateEnergy}) could
have been also obtained by assuming that the population of all states in the
same energy shell is equal,
\begin{equation}
\label{ergo:2} \langle {\bf n}|\rho|{\bf n}\rangle = p(E_{\bf n}),
\end{equation}
thus defining $p(E)$ as the population of a state at energy $E$, such that
(c.f. Eq.~(\ref{P:E}))
\begin{equation}
\label{p(E)} P(E)=p(E)g(E).
\end{equation}
Equation~(\ref{ergo:2}) implies (\ref{ergo:1}) and is an even stronger
assumption.
It corresponds to assuming that the system behaves {\it ergodically}, i.e. that 
the populations of the states in an energy shell equalize faster than the 
average quantities of the system evolve in time.
It leads to the same rate equation (\ref{rateEnergy}), but it is a more natural
assumption in the last stages of the evolution, when the system tends
asymptotically to the steady state.

In general, (\ref{rateEnergy}) describes the dynamics of cooling to the
extent that assumption (\ref{ergo:1}) is valid, i.e. when the average
coupling between an eigenstate of the mechanical energy and the states of an
energy shell is a smooth function of their respective energy. This is
true when the average coupling of each state of one energy shell to the states of
another energy shell is of the same order of magnitude. Outside the
Lamb-Dicke regime, given the oscillatory behaviour of the Franck-Condon
coefficients, this holds if each state couples appreciably to more than
one state of the other energy shell, i.e. for $D(E)\gg 1$ and $\gamma$
sufficiently large. In the Lamb-Dicke regime analogous considerations can be
applied, and they will be discussed in detail in the next section.

The ergodic regime can also be justified by the existence of a physical
process whose main effect is to thermalize the states within one energy shell
and which acts on a shorter time scale than the energy-changing processes.
This assumption is at the basis of treatments in the kinetic theory of
quantum gases \cite{ErgoBEC}, where the interatomic collisions lead to
thermalization, and the gas can be considered in a thermal quasi-equilibrium
distribution on the time scale in which it is cooled. The assumption of rapid
thermalization is also central to an earlier study of laser cooling of
Coulomb clusters \cite{Javanainen}: There the effect of mode-mode coupling
(anharmonicity) has been proposed as a possible agent, which does not
explicitly appear in the equations but justifies condition (\ref{ergo:1}).
Direct evidence of this effect has been found in numerical studies in
\cite{EPJD} for the case of exact degeneracy between the modes frequencies.

\subsubsection{On the laser intensity distribution over the crystal}

\noindent In deriving (\ref{rateEnergy}) we have assumed that all ions are
uniformly driven. We discuss now the case in which only a subset $\{j\}$ of
ions in the crystal are illuminated. For simplicity, let us assume in
Eq.~(\ref{rate2}) that $\Omega_l=\Omega$ for $l\in\{j\}$, and $\Omega_l=0$
otherwise. Due to the geometry of the crystal, the coupling of each ion to a
mode is a function of the mode and of the ion's position in the crystal [see
Eq.~(\ref{LambDicke})]. Some positions in the chain are more strongly coupled
to a certain set of modes, and some positions are even decoupled from certain
modes \cite{EPJD}. Then, the ergodic assumption is valid provided a
suitable set of ions is driven, such that the coupling of each mode of
collective motion to the radiation is of the same order of magnitude. In that
case, the ergodic assumption, Eq.~(\ref{ergo:1}), can be expressed as:
\begin{equation}
\label{ergo:M} \frac{1}{M} \sum_{\{j\}} {\sum_{\bf k}}^{\prime}
|\langle {\bf k}| \text{e}^{-ikz_j}|{\bf n}\rangle |^{2} \approx
D(E_{\bf k}) Q^{(k)}(E_{\bf n},E_{\bf k}),
\end{equation}
where $M$ is the number of driven ions ($M\le N$). Taking into account this
more general case, the rate equation (\ref{rateEnergy}) has the form:
\begin{eqnarray}
\frac {{\rm d}}{{\rm d}t} P(E,t) &=& -\gamma \frac{\Omega^{2}}{4} M
\int_0^{\infty}\text{d} E_1 \frac {g(E_1) Q^{(k_z)}(E,E_1)}
{\left[ (E_1-E)/\hbar -\delta \right] ^{2}+\gamma ^{2}/4} P(E,t)
\nonumber \\
&+& \gamma \frac{\Omega^{2}}{4}M \int_{-1}^{1} \text{d}(\cos\theta)
{\mathcal N}(\cos\theta) \int_0^{\infty} \text{d} E_1
\int_0^{\infty} \text{d} E_2 \frac {g(E) Q^{(k\cos\theta)}(E_1,E)
g(E_1) Q^{(k_z)}(E_2,E_1)} {\left[ (E_1-E_2)/\hbar-\delta \right]
^{2}+\gamma ^{2}/4} P(E_2,t). \label{rateEnergy:1}
\end{eqnarray}
Finally, note that the number of driven ions, $M$, appears in
(\ref{rateEnergy}) and (\ref{rateEnergy:1}) as an overall scaling factor,
making the optical pumping rate of the crystal $M$ times that of a single
ion. In fact, it has been shown earlier that below saturation the ions behave
as independent scatterers, and their contributions to cooling simply add up
\cite{PRAIbk,EPJD}.

\section{The semiclassical limit and the Lamb-Dicke limit}

\noindent In this section, starting from Eq.~(\ref{rateEnergy}) or
(\ref{rateEnergy:1}) we derive the limit where the motion can be treated
classically (semiclassical limit), and obtain a Fokker-Planck-equation for the
energy which is analytically solvable. We compare our result with the
well-known treatments of \cite{JavaSten} and \cite{Javanainen} and find
full agreement with their theoretical predictions. In the second part of this
section, starting from (\ref{no_neglect}) we consider the Lamb-Dicke limit,
and discuss the conditions under which an equation for the motional energy of
the type of (\ref{rateEnergy}) can be derived for describing cooling of $N$
ions.

For clarity in the derivation, we rewrite Eq.~(\ref{rateEnergy}) as:
\begin{eqnarray}
g(E)\frac{\rm d}{{\rm d}t}p(E,t) &=& -
g(E)p(E,t)\int_0^{\infty}\text{d} E_1 f_E^{(k_z)}(E_1)L(E_1-E)
\nonumber \\
&+&g(E)\int_{-1}^{1}\text{d}(\cos\theta){\mathcal
N}(\cos\theta)\int_0^{\infty}\text{d} E_1 \int_0^{\infty}\text{d}
E_2 f^{(k\cos\theta)}_E(E_1)f_{E_1}^{(k_z)}(E_2)
L(E_1-E_2)p(E_2,t). \label{rate7}
\end{eqnarray}
Here, $p(E)$ is the population of a state of energy $E$ defined in
(\ref{ergo:2}), and $L(E_1-E)$ is the Lorentzian distribution,
\begin{equation}
L(E_1-E)=\frac{M\Omega^2\gamma}{4\left[ (E_1-E)/\hbar -\delta \right]
^{2}+\gamma ^{2}},
\end{equation}
while $f^{(k)}_E(E^{\prime})$ is defined through:
\begin{equation}
g(E^{\prime})Q^{(k)}(E,E^{\prime}) = f_E^{(k)}(E^{\prime}).
\label{equivalence}
\end{equation}
From this definition, using (\ref{ergo:1}) and (\ref{Center1}-\ref{width}),
it can be verified that $f^{(k)}_E(E^{\prime})$ fulfills the relations:
\begin{eqnarray}
\label{Prop}
&&\int_0^{\infty}dE^{\prime} f_E^{(k\cos\theta)}(E^{\prime})=1,\\
\label{Prop1}
&&\int_0^{\infty}dE^{\prime} (E^{\prime}-E)
f_E^{(k\cos\theta)}(E^{\prime}) = \hbar\omega_R\cos^2\theta, \\
\label{Prop2} &&\int_0^{\infty}dE^{\prime} (E^{\prime}-E)^2
f_E^{(k\cos\theta)}(E^{\prime}) = (\hbar\omega_R\cos^2\theta)^2 +
\hbar\omega_R\cos^2\theta \frac{2E}{N}.
\end{eqnarray}
The analogous relations follow for $f_E^{(k_z)}(E^{\prime})$ when $\theta$ is
replaced by $\theta_0$, where $k_z = k\cos\theta_0$. \\
Equations~(\ref{Prop1}) and (\ref{Prop2}) are equivalent to
(\ref{Center1}) and (\ref{width}) in the approximation of a smooth energy
scale.

\subsection{Derivation of a Fokker-Planck equation for the energy}

\noindent We first consider the limit in which the recoil frequency is much
smaller than the linewidth, $\omega_R\ll \gamma$. In this limit, while
$f_E^{(k)}(E^{\prime})$ varies on the scale $\hbar\omega_R$ (see
Eqs.~(\ref{Prop1}-\ref{Prop2})), the variation of the population $p(E)$ on
the energy scale $\hbar\omega_R$ is small, i.e.
\begin{equation}
\label{FokkerApprox}
\hbar\omega_R |\frac{\partial}{\partial E}p(E,t)|\ll p(E,t).
\end{equation}
We now expand Eq.~(\ref{rate7}) around $E$ up to first order in the
parameter
$\omega_R/\gamma$ and, using (\ref{Prop}-\ref{Prop2}), we obtain:
\begin{eqnarray}
\label{rate8} \frac{\rm d}{{\rm d}t}g(E)p(E,t)
&=& - 2 \omega_R\cos^2\theta_0  L^{\prime}(0)g(E) p(E,t)\\
&+& \omega_R
\left[(\alpha+\cos^2\theta_0)L(0)-2\frac{E}{N}\cos^2\theta_0
L^{\prime}(0)\right]g(E)p^{\prime}(E,t)
\nonumber\\
&+& \omega_R \frac{E}{N}(\alpha+\cos^2\theta_0)L(0)
g(E)p^{\prime\prime}(E,t). \nonumber
\end{eqnarray}
Here,
\begin{equation}
L^{\prime}(0)=\left[\frac{\rm d}{{\rm d}x}
L(x)\right]_{x=0}=\frac{4M\Omega^2\gamma\delta}{(4\delta^2+\gamma^2)^2},
\label{L1}
\end{equation}
and
\begin{equation}
\label{alpha} \alpha=\int {\rm d}(\cos\theta) \cos^2\theta {\mathcal
N}(\cos\theta)~.
\end{equation}
We rescale the time as
\begin{equation}
\label{tau} \tau=\omega_R (\cos^2\theta_0+\alpha)L(0)~t
\end{equation}
and define
\begin{equation}
C = 2~ \frac {\cos^2\theta_0} {(\cos^2\theta_0+\alpha)}~ \frac
{L^{\prime}(0)} {L(0)}.
\end{equation}
Note that that with definition (\ref{L1}), $C<0$ for red detunings $\delta<0$
($\omega_L<\omega_0$). Eq.~(\ref{rate8}) can now be written as:
\begin{equation}
\frac{\rm d}{{\rm d}\tau} g(E)p(E,\tau) =-\frac{C}{N}\left[N
g(E)p(E,\tau) + Eg(E)p^{\prime}(E,\tau) \right] +
\frac{1}{N}\left[Ng(E)p^{\prime}(E,\tau) +
Eg(E)p^{\prime\prime}(E,\tau)\right].
\end{equation}
Using the energy dependence of the smoothed density of states $g(E)$ as given
in (\ref{smoothDensity}), and (\ref{p(E)}), we get a Fokker-Planck Equation of the
form:
\begin{equation}
\label{rate8:1} \frac{\rm d}{{\rm d}\tau}P(E,\tau) =
-\frac{\partial}{\partial E}\left[A(E)P(E,\tau)\right]
+\frac{1}{2}\frac{\partial^2}{\partial
E^2}\left[B(E)P(E,\tau)\right],
\end{equation}
where
\begin{eqnarray}
A(E)&=&1+\frac{C}{N}E,\nonumber\\
B(E)&=&\frac{2}{N}E.\nonumber
\end{eqnarray}
Eq.~(\ref{rate8:1}) can be easily solved. In the following, we calculate the
steady state solution and the time evolution of the system, and compare the
result with the existing treatments evaluated in the limit of one ion ($N=1$).

\subsubsection{Steady State}

\noindent The steady state distribution $P_0(E)$ satisfies the equation
$\frac{\rm d}{{\rm d}\tau}P_0(E)=0$. Thus, it is solution of the differential
equation:
\begin{equation}
\label{SteadyCL} \frac{\partial}{\partial E}\left[B(E)P_0(E)\right] =
2A(E)P_0(E),
\end{equation}
(where the integration constant has been set to 0 for a convergent solution)
and has the form
\begin{equation}
\label{SteadyFP} P_0(E) = FE^{N-1}\exp \left( CE \right),
\end{equation}
with normalization constant $F$. For red detunings ($C<0$) and if the wave
vector of the cooling laser has a component along the motional axis
($\cos\theta_0\neq 0$) the integral of (\ref{SteadyFP}) over the energies
converges. Then, the value of $F$ is found from $\int_0^{\infty} {\rm d}E
P_0(E)=1$, yielding:
\begin{equation}
F=\frac{|C|^N}{\Gamma(N)}.
\end{equation}
Using these results, the steady state energy has the form:
\begin{eqnarray}
\langle E\rangle &=&\int_0^{\infty} {\rm d}E E P_0(E)=\frac{N}{|C|}
\nonumber\\
                 &=&N\gamma~ \frac{\alpha+\cos^2\theta_0}{4\cos^2\theta_0}
\left(\frac{\gamma}{2|\delta|}+\frac{2|\delta|}{\gamma}\right).
\label{Esteady}
\end{eqnarray}

Eq.~(\ref{Esteady}) contains the dependence of the cooling limit on the angle
$\theta_0$ between the cooling laser beam and the direction of the motion.
The final energy is minimal for $\cos\theta_0=1$, i.e. when the laser
propagates parallel to the trap axis, and it diverges for $\cos\theta_0=0$,
when the laser is orthogonal to the trap axis and there is thus no laser cooling.

The minimum of the final energy {\it vs.} detuning is reached for
$\delta=-\gamma/2$, as in the case of one ion (see for example
\cite{Stenholm}). Inserting into (\ref{Esteady}) $N=1$, $\alpha=1/3$ (which
corresponds to spatially isotropic spontaneous emission), and
$\cos\theta_0=1$, we find the same result as Javanainen and Stenholm in their
semiclassical expansion for one ion \cite{JavaSten}. Hence, the final energy
for $N$ ions is $N$ times the steady state energy achieved by Doppler cooling
of one ion. This general result has also been found earlier in special cases,
such as a Coulomb cluster in the Lamb-Dicke regime \cite{Javanainen}, and a
two-ion crystal treated in Ref.~\cite{Giovi} by extending the method of
\cite{JavaSten}.

\subsubsection{Time Evolution}

\noindent The steady state solution suggests to use the following ansatz for
solving Eq.~(\ref{rate8:1}):
\begin{equation}
\label{Ansatz} P(E,t) = F(t) E^{N-1} \exp \left( -\frac{E}{U(t)}
\right),
\end{equation}
where $U(t)$ is a positive function of time and $F(t)$ is a
normalization factor, such that at any instant $t$ the relation
$\int_0^{\infty}{\rm d}E F(t) P(E,t)=1$ is fulfilled, i.e.
\begin{equation}
F(t)=\frac{1}{\Gamma(N)U(t)^N}.
\end{equation}
This ansatz corresponds to assuming that the distribution $P(E,t)$
is always thermal with average energy
\begin{equation}
\label{AnsUsed} \langle E(t)\rangle=NU(t).
\end{equation}
Substituting (\ref{Ansatz}) into Eq.~(\ref{rate8:1}) and using
(\ref{tau}) we get a differential equation for $U$:
\begin{equation}
\label{timeAnsatz} \frac{{\rm d}U}{{\rm d}\tau} \frac{1}{U} \left(
\frac{E}{U}-N \right) = \frac{1}{N} \left( C+\frac{1}{U} \right)
\left( \frac{E}{U}-N \right).
\end{equation}
Since this equation must be fulfilled for all values of $E$, we
find:
\begin{equation}
\frac{{\rm d}U}{{\rm d}\tau} = \frac{1}{N} \left( CU+1 \right),
\end{equation}
which has the solution:
\begin{equation}
\label{SoluTime} U(t)=U_0\exp\left(\frac{2\omega_R \cos^2\theta_0
L^{\prime}(0)}{N}~t\right)+\frac{1}{C},
\end{equation}
where $NU_0=\langle E(0)\rangle$ is the initial energy. For $t\to\infty$ we
recover from Eq.~(\ref{SoluTime}) the steady state solution (\ref{SteadyFP}).
The rate at which the steady state is reached is:
\begin{equation}
\label{SoluRate} \Gamma_{\rm cool}=\frac{2\omega_R \cos^2\theta_0
|L^{\prime}(0)|}{N} = \frac{M}{N}
\frac{8\omega_R\cos^2\theta_0\Omega^2\gamma|\delta|}{(4\delta^2+\gamma^2)^2},
\end{equation}
where we have used (\ref{L1}). Thus the cooling rate increases linearly with
the number of driven ions $M$. It is largest if all ions are driven, as
expected.

Since the results derived so far agree precisely with those found earlier in
specific cases, the general procedure which lead us from the quantum
mechanical equations to the rate equation for the energy can be considered
the common basis which underlies and unifies these earlier treatments.
Furthermore, we have shown that the final energy of an $N$-ion crystal is $N$
times that of a single motional mode, while the cooling rate is $M$ times
that for the one-dimensional motion of a single ion, $M$ being the number of
ions driven by the laser.

A final remark should be made on the assumptions leading to (\ref{rate8:1}).
In deriving the Fokker-Planck equation we have assumed ergodicity and
Eq.~(\ref{FokkerApprox}), i.e. that $p(E)$ varies negligibly on the recoil
energy scale. The latter condition corresponds to the second-order expansion
in $\hbar$ of Ref.~\cite{JavaSten}. In that work the derivation of a
Fokker-Planck equation was based on the limit of overdamped oscillation,
$\gamma\gg\nu$, in order to adiabatically eliminate the excited state from
the equations. Our derivation, however, does not necessarily imply this
limit. Only in the case of $N=1$ ion, which is not the main focus of our
study, we must have $\Delta E\gg \hbar\nu$ in order to fulfill the condition
of a large density of states, $D(E)\gg 1$, and thus for this special case the
overdamped oscillator limit is a requirement for the validity of (\ref{rate8:1}).

\subsection{Lamb-Dicke regime}

\noindent When the atomic motion is well localized with respect to the laser
wavelength (Lamb-Dicke regime), the Franck-Condon coefficients in (\ref{fc})
can be approximated by their first order expansion in the parameter $\langle
(\vec{k}\cdot\vec{x})^2 \rangle$. For a single ion excited below saturation,
the dynamics are described by a rate equation for the populations of the
states with vibrational number $n$, which can be analytically solved
\cite{Stenholm,Lindberg}. In this form, since $n$ is proportional to the mechanical
energy of the ion, the equation for cooling
is an equation for the energy. For many ions,
a set of rate equations can also be derived which describe cooling of each
mode and which are decoupled, since a simultaneous change of the
energy of two or more modes by scattering of a photon is of higher order in
the Lamb-Dicke parameter (and thus takes place on a longer time scale)
\cite{PRAIbk,EPJD}.

Yet we show in this section that under some conditions we can derive a single
equation for the total energy of many ions. Given the analytical simplicity
of the model, this example is also instructive in order to see the limits of
validity of the ergodic equation.

Let us consider Eqs.~(\ref{no_neglect}) in the Lamb-Dicke regime
\cite{LDRmany}. Here the reduction of this set of equations into
the rate equations (\ref{rate2}) is possible without assumption
(\ref{RandomPhase}), since the terms
(\ref{line2}), (\ref{line4}) are negligible either because they
are rapidly oscillating or because their coupling to the
populations is of higher order in the Lamb-Dicke parameter
\cite{PRAIbk}. In first order in the parameter
$\omega_R/\nu_{\beta}$, with $\beta=1,...,N$, the modulus square
of the Franck-Condon coefficient has the form:
\begin{eqnarray}
\label{Eq4:exp_eta} |\langle {\bf n}|\text{e}^{ikz_j}|{\bf
k}\rangle |^{2}
&\approx&\prod_{\alpha=1}^N\delta_{n_{\alpha},k_{\alpha}}
\left(1-\sum_{\beta=1}^N{\eta^{\beta}_j}^2(2n_{\beta}+1)\right)\\
&+&\sum_{\beta=1}^N\prod_{\alpha=1,\alpha\neq
\beta}^N\delta_{n_{\alpha},k_{\alpha}} \left[
{\eta^{\beta}_j}^2(n_{\beta}+1)\delta_{k_{\beta},n_{\beta}+1}+
 {\eta^{\beta}_j}^2 n_{\beta}\delta_{k_{\beta},n_{\beta}-1}\right].
\nonumber
\end{eqnarray}
Thus, given the initial state of motion $|{\bf n}\rangle =
|n_1,...,n_N\rangle$, scattering of a photon involves three
possible transitions: (i) the so-called carrier transition, of
zero order in the Lamb-Dicke parameter, with final state $|{\bf
k}\rangle = |{\bf n}\rangle$; (ii) the red-sideband transitions,
where $|{\bf k}\rangle = |n_1,...,n_{\beta}-1,...,n_N\rangle$,
and (iii) the blue-sideband transitions, where $|{\bf k}\rangle =
|n_1,...,n_{\beta}+1,...,n_N\rangle$. The cases (ii) and (iii)
are of second order in the parameter $\eta_{\beta}$.

When we substitute (\ref{Eq4:exp_eta}) into Eq.~(\ref{rate2}),
we obtain:
\begin{eqnarray}
\label{LDLrate1} \frac{\rm d}{{\rm d}t}\langle {\bf n}|\rho|{\bf
n}\rangle & = & - \langle {\bf n}|\rho|{\bf
n}\rangle\omega_R\cos^2\theta_0 \sum_{\beta=1}^N
\left[(n_{\beta}+1)L(\nu_{\beta})/\nu_{\beta}+
n_{\beta}L(-\nu_{\beta})/\nu_{\beta}\right]\\
& - & \langle {\bf n}|\rho|{\bf n}\rangle \omega_R
\sum_{\beta=1}^N\alpha(2n_{\beta}+1)L(0)/\nu_{\beta}
\nonumber\\
& + & \sum_{\beta=1}^N  \langle {\bf n} +{\bf 1}_{\beta}|\rho
|{\bf n}+{\bf 1}_{\beta}\rangle
\frac{\omega_R}{\nu_{\beta}} (n_{\beta}+1)
\left[\cos^2\theta_0 L(-\nu_{\beta})+\alpha L(0)\right]
\nonumber\\
& + & \sum_{\beta=1}^N \langle {\bf n}-{\bf 1}_{\beta}|\rho|
{\bf n}-{\bf 1}_{\beta}\rangle
\frac{\omega_R}{\nu_{\beta}} n_{\beta} \left[\cos^2\theta_0
L(\nu_{\beta})+\alpha L(0)\right],
\nonumber
\end{eqnarray}
where have used $\sum_j{\eta^{\beta}_j}^2 = \omega_R/\nu_{\beta}$, and the
vector ${\bf 1}_{\beta}$ is defined in the $N$-dimensional Hilbert space
through $({\bf 1}_{\beta})_{\alpha} =\delta_{\alpha,\beta}$ for $\alpha=1,...,N$.

Eq.~(\ref{LDLrate1}) is linear in the vibrational numbers $n_{\beta}$ of the
single modes, which are weighted by the factors $\omega_R L(\nu_{\beta}) /
\nu_{\beta}$. Only when these weights are of approximately equal magnitude,
a hypothesis similar to (\ref{ergo:1}), which allows summation over
states of equal energy, can be applied. This is true when
$\nu_{\beta}\ll\gamma$ for all modes $\beta$ because in this limit the
Lorentzian atomic resonance curve varies very slowly over all red (blue)
sidebands of the modes. Then, by expanding the terms in (\ref{LDLrate1}) at
the second order in the parameter $\nu_{\beta}/\gamma$, we obtain a
Fokker-Planck equation of the same form as (\ref{rate8:1}). In that sense,
the condition $\nu_{\beta}\ll\gamma$ can be considered a requirement for
deriving an ergodic equation in the Lamb-Dicke regime.

Alternatively, in \cite{Javanainen} a single equation for the total motional
energy in the Lamb-Dicke regime has been justified by assuming that all modes
thermalize on a faster time-scale and thus imposing
\begin{equation}
\label{ergo:3} \langle
n_{\beta}\rangle=\frac{1}{N}\frac{E}{\hbar\nu_{\beta}}~.
\end{equation}
Application of (\ref{ergo:3}) to (\ref{LDLrate1}) yields an equation for the
total mechanical energy which can be solved analytically. In our treatment
above, in the case where we find a Fokker-Planck equation, condition
(\ref{ergo:3}) is also fulfilled, but there it is a consequence of the average 
coupling among energy shells due to light scattering, rather than an extra assumption.

\subsection{Discussion}

\noindent The Fokker-Planck equation (\ref{rate8:1}) has been derived in two
different ways: in the semiclassical case by starting from the ergodic
equation (\ref{rateEnergy}), and in the Lamb-Dicke regime by starting from
equation (\ref{rate2}). Both derivations rely on the limit
$\gamma\gg\nu,\omega_R$, while in the Lamb-Dicke regime there is the
additional constraint that $\omega_R\ll\nu$.

Let us now summarize the solutions of (\ref{rate8:1}). The form of the
solution is the same as the one obtained for cooling of one ion, whereby here
the number $N$ of ions appears in the steady-state energy (\ref{Esteady}) as
a scaling factor, and the cooling rate (\ref{SoluRate}) scales as the ratio
$M/N$ where $M$ is the number of driven ions. This result supports the
conclusion drawn in \cite{Javanainen}, that the cluster is cooled like a
single ion. It is instructive, at this regard,
to compare two particular cases which exhibit
analogies: The axial motion of two ions in a trap with collective modes at
frequencies $\nu_1$, $\nu_2$, and the motion of a single ion in a
two-dimensional harmonic oscillator potential with the same frequencies,
$\nu_1$ and $\nu_2$, on the two axes.
Neglecting for the moment the different
masses, the mechanical Hamiltonians of the two systems are equivalent. The
laser--ion interaction terms are also equivalent, provided that one of the
two ions in the chain is driven, and that the Lamb-Dicke parameter for each
mode of the chain is the same as the Lamb-Dicke parameter for each axis of
the trap of the single ion. This latter condition can be fulfilled by
choosing, in the single-ion case, the proper angle of propagation of the
laser in the two-dimensional plane.

The main difference between the two cases arises in the spontaneous emission:
in the two--dimensional one--ion case, the photon is scattered at a random
angle in the plane and the ratio between the projections of the recoil energy
on each axis can vary. In contrast, in the two--ion case, the scattering
angle is the same for both modes, hence the average energy transfer to the
two modes has a fixed ratio in any scattering event. This difference appears
as a the geometrical factor in the dynamical equations, as well as in the
expression for the final energy (\ref{Esteady}) \cite{Itano}. From that
consideration we expect that the generalization of our treatment to a
3-dimensional Coulomb crystal will change the one-dimensional results only by
numerical factors representing the different geometry of the problem.

\section{Evaluation of the semiclassical Ergodic Equation}

\noindent So far we have assumed a grid of energy $\Delta E$, which defines
the coarse-graining on which the relevant system properties for cooling
are defined. The grid $\Delta E$ has been chosen according to the condition
$\hbar\gamma \gg \Delta E$, and an equation for the energy has been derived
for the situation that the number of states at energy $E$ is $D(E)\gg 1$.
Thus, $\Delta E$ represents a limited resolution which allows us to average
over many quantum states. We now discuss the limit $\Delta E\gg \hbar\nu_N$.
This case is expected to correspond to the classical limit, since the
resolution is such that all details of the quantum spectrum are averaged out.
We will derive the classical limit starting from the
average Franck-Condon coefficients $Q^{(k)}(E,E^{\prime})$ in
(\ref{averageFC}), from which we obtain an explicit form for the ergodic
equation (\ref{rateEnergy}).

Let us consider the sum in Eq.~(\ref{averageFC}). Using the property of the
trace, we can write:
\begin{eqnarray}
\frac{1}{N}\sum_j{\sum_{\bf k}}^{\prime}{\sum_{\bf n}}^{\prime} \langle {\bf
k}| \text{e}^{-ikz_j}|{\bf n}\rangle \langle {\bf n}| \text{e}^{ikz_j}|{\bf
k}\rangle &=& \frac{1}{N}\sum_j {\sum_{\bf k}}^{\prime}{\sum_{\bf
n}}^{\prime} \text{Tr}\left\{|{\bf k}\rangle \langle {\bf k}|
\text{e}^{-ikz_j}|{\bf
n}\rangle \langle {\bf n}| \text{e}^{ikz_j} \right\} \nonumber \\
&{\equiv}& C(E_{\bf k},E_{\bf n})~, \label{Trace1}
\end{eqnarray}
thus defining $C(E,E^{\prime}) = Q^{(k)}(E,E^{\prime}) D(E) D(E^{\prime})$.
The summation over the states belonging to the shell of energy $E$ can be
re-expressed as:
\begin{equation}
{\sum_{\bf k}}^{\prime}|{\bf k}\rangle \langle {\bf k}| = \int_{{\cal
F}(E_{\bf k})} {\rm d}E\delta(E-\hat{H}_{\rm mec})\sum_{{\bf k}} |{\bf
k}\rangle \langle {\bf k}|,
\end{equation}
and Eq.~(\ref{Trace1}) aquires the form:
\begin{eqnarray}
C(E_{\bf k},E_{\bf n}) = \frac{1}{N} \sum_j\text{Tr}\left\{ \int_{{\cal
F}(E_{\bf k})} \text{d}E \int_{{\cal F}(E_{\bf n})}
\text{d}E^{\prime}\delta(E-\hat{H}_{\rm mec}) \text{e}^{ikz_j}
\delta(E^{\prime}-\hat{H}_{\rm
mec})\text{e}^{-ikz_j}\right\}. \nonumber \\
\label{Trace2}
\end{eqnarray}
Using the Fourier transform of the $\delta$-function, we get:
\begin{eqnarray}
\int_{{\cal F}(E_{\bf k})} {\rm d}E\delta(E-\hat{H}_{\rm mec}) = \int_{{\cal
F}(E_{\bf k})} {\rm d}E \frac{1}{2\pi\hbar} \int_{-\infty}^{\infty} {\rm
d}\tau\mbox{e}^{i(E-\hat{H}_{\rm mec})\tau/\hbar} = \frac{1}{2\pi}
\int_{-\infty}^{\infty} {\rm d}\tau \frac{\sin(\Delta E\tau/\hbar)}{\tau}
\mbox{e}^{i(E_{\bf k}-\hat{H}_{\rm mec})\tau/\hbar}, \label{Fourier}
\end{eqnarray}
such that Eq.~(\ref{Trace2}) can be rewritten as:
\begin{equation}
C(E,E^{\prime}) =\frac{1}{N}\sum_j\frac{1}{4\pi^2}
\int_{-\infty}^{\infty}{\rm d}\tau\frac{\sin(\Delta
E\tau/\hbar)}{\tau} \int_{-\infty}^{\infty}{\rm d}\tau^{\prime}
\frac{\sin(\Delta
E\tau^{\prime}/\hbar)}{\tau^{\prime}}A_j(\tau,\tau^{\prime}),
\label{Trace3}
\end{equation}
where we have defined:
\begin{equation}
A_j(\tau,\tau^{\prime})={\rm Tr}\left\{\mbox{e}^{i(E-\hat{H}_{\rm
mec})\tau/\hbar}\mbox{e}^{ikz_j}
\mbox{e}^{i(E^{\prime}-\hat{H}_{\rm
mec})\tau^{\prime}/\hbar}\mbox{e}^{-ikz_j}\right\}. \label{Trace3a}
\end{equation}
By using coherent states $|\alpha_{\beta}\rangle$ to calculate the trace,
using (\ref{quant:q}), (\ref{kick}), and the properties of coherent states
\cite{Gardiner}, Eq.~(\ref{Trace3a}) takes the form:
\begin{eqnarray}
A_j(\tau,\tau^{\prime})
& =      &\prod_{\beta=1}^N\int\frac{{\rm
d}^2\alpha_{\beta}}{\pi^{2N}}
\exp\left[-i\frac{\nu_{\beta}}{2}(\tau+\tau^{\prime})\right]
\exp\left[-|\alpha_{\beta}|^2{\rm
e}^{-i\nu_{\beta}\tau^{\prime}}(1-{\rm
e}^{-i\nu_{\beta}\tau})\right]
\nonumber\\
&\cdot   &
\exp\left[-\left(|\alpha_{\beta}|^2+{\eta_j^{\beta}}^2+i\eta_j^{\beta}
(\alpha^*_{\beta}{\rm e}^{-i\nu_{\beta}\tau}-\alpha_{\beta})\right)
(1-{\rm e}^{-i\nu_{\beta}\tau^{\prime}})\right].
\label{Trace4:N:2}
\end{eqnarray}
In Eq.~(\ref{Trace3}), the integrand is the product of
$A_j(\tau,\tau^{\prime})$ and the window function $\sin(x)/x$ of width
$\hbar/\Delta E$. According to Eq.~(\ref{Trace4:N:2}),
$A_j(\tau,\tau^{\prime})$ depends on $\tau, \tau^{\prime}$ only through ${\rm
e}^{i\nu_{\beta}\tau}$ and ${\rm e}^{i\nu_{\beta}\tau^{\prime}}$, and thus
$\tau$ is scaled by the mode frequencies. 

\subsection{Derivation of the classical limit}

\noindent We now consider the limit $E,\Delta E\gg \hbar \nu_{\beta}$ for all
$\beta$. This corresponds to the situation $\langle n_{\beta} \rangle \gg 1$,
e.g. before the cooling limit is reached, and in general to the regime 
$\gamma\gg\nu_{\beta}$. In this limit we can expand the exponentials in
(\ref{Trace4:N:2}) in the parameter $\hbar\nu/\Delta E$. Up to first order,
Eq.~(\ref{Trace3}) is equivalent to:
\begin{eqnarray}
C(E,E^{\prime}) &\approx &\left(\frac{\Delta
E}{\hbar}\right)^2
\prod_{\beta=1}^N\int\frac{{\rm d}^2\alpha_{\beta}}{\pi^{2N}}
\left[\frac{1}{2\pi} \int_{-\infty}^{\infty}{\rm
d}\tau\frac{\sin(\Delta E\tau/\hbar)}{\Delta E\tau/\hbar}
\mbox{e}^{i(E/\hbar-
\nu_{\beta}(|\alpha_{\beta}|^2+1/2))\tau}\right]\nonumber\\
&\cdot   & \left[\frac{1}{2\pi}\int_{-\infty}^{\infty}{\rm
d}\tau^{\prime} \frac{\sin(\Delta E\tau^{\prime}/\hbar)}{\Delta
E\tau^{\prime}/\hbar}
\mbox{e}^{i(E^{\prime}/\hbar-\nu_{\beta}(|\alpha_{\beta}|^2+1/2-2\eta^{\beta}
{\rm Im}(\alpha_{\beta}) -\eta^{\beta 2}))\tau^{\prime}}\right].
\label{Trace4:N:3}
\end{eqnarray}
This result can be interpreted physically, considering that $\tau$ is the
Fourier conjugate of the energy $E/\hbar$ and has the dimension of a time.
Thus, Eq.~(\ref{Trace3a}) can be seen as the overlap between the state $|{\bf
\alpha}\rangle$ and the state $|{\bf
\alpha}^{\prime}\rangle$, which corresponds to a displacement of $|{\bf
\alpha}\rangle$, followed by free evolution 
for a time $\tau$, then by a displacement back, and finally by 
free evolution for a time
$\tau^{\prime}$. Integral (\ref{Trace3}) sums over all intervals of times
$\tau$, $\tau^{\prime}$, for all trajectories in the neighbourhood of the energy shells $E$ 
at $\tau$ and $E^{\prime}$ at $\tau^{\prime}$, 
weighted by the time window of resolution $\hbar/\Delta E$. For
$\tau=\tau^{\prime}=0$ the overlap is maximum, since the state is unchanged.
Although the overlap shows a certain periodicity, e.g. for $N=1$ and
$\tau,\tau^{\prime}$ multiples of $2\pi/\nu$, $A(\tau,\tau)=A(0,0)$, for
$\Delta E\gg \hbar\nu$ these recurrence times fall out of the window function
interval. Therefore the only appreciable contribution to the integral comes
from $\tau,\tau^{\prime}\sim0$. In other words, the classical limit corresponds
to large energy uncertainty, such that only the short time evolution of
the wave packet contributes appreciably to the integral, since only in this
limit the trajectories are phase coherent.

Eq.~(\ref{Trace4:N:3}) can now be rewritten as an integral in the classical
phase space using the definition of coherent states
\begin{equation}
\alpha_{\beta}=i\bar{p}_{\beta}\sqrt{\frac{\hbar}{2m\nu_{\beta}}}
+\bar{q}_{\beta}
\sqrt{\frac{m\nu_{\beta}}{\hbar}},
\end{equation}
where $\bar{q}_{\beta},\bar{p}_{\beta}$ are real numbers. Integrating over
$\tau,\tau^{\prime}$ yields:
\begin{eqnarray}
\label{ClassicTrace} C(E,E^{\prime}) &=& \int_{{\cal F}(E)} \text{d}E_1
\int_{{\cal F}(E^{\prime})} \text{d}E_2 \int \frac{{\rm d}\bar{q}_1...{\rm
d}\bar{q}_N{\rm d}\bar{p}_1...{\rm d}\bar{p}_N}{h^N}\\
&\times& \delta \left( E_1 -\sum_{\beta=1}^N \frac{\bar{p}_{\beta}^2}{2m} -
V(\bar{q}_1,...,\bar{q}_N) \right)\nonumber\\
&\times& \delta \left( E_2 -\frac{(\bar{p}_1-\hbar k)^2}{2m} -
\sum_{\beta=2}^N\frac{\bar{p}_{\beta}^2}{2m}
-V(\bar{q}_1,...,\bar{q}_N)\right).\nonumber
\end{eqnarray}
Integrating in phase space [see Appendix C], and using the relation
$f_E(E^{\prime}) = C(E,E^{\prime})/(D(E^{\prime})\Delta E)$ we obtain:
\begin{equation}
f_E(E^{\prime})=\frac{\Gamma(N)/\sqrt{4\hbar\omega_R
E}}{\sqrt{\pi}\Gamma(N-\frac{1}{2})}
\left[1-\left(\frac{E^{\prime}-E-\hbar\omega_R}{\sqrt{4\hbar\omega_R
E}}\right)^2 \right]^{N-3/2}. \label{SemiAns}
\end{equation}
The function $f_E(E^{\prime})$ is real, and thus well-defined, on the interval
of energies $[E+\hbar\omega_R-\sqrt{4\hbar\omega_R E},
E+\hbar\omega_R+\sqrt{4\hbar\omega_R E}]$, and it is normalized with respect
to $E^{\prime}$. In Fig.~3 we plot Eq.~(\ref{SemiAns}) for
$N=1,10,100$. For $N=1$ it has the well-known form of the classical
momentum distribution of a harmonic oscillator at a given energy $E$: In fact
in this case $E^{\prime}=E+\hbar\omega_R+\hbar k p/m$, {\it i.e.} the
probability for the oscillator to have final energy $E^{\prime}$ after the
scattering is equal to the probability that the oscillator, at energy $E$,
has momentum $p$ when the scattering event occurs. For higher $N$ the
distribution peaks more and more narrowly around $E+\hbar\omega_R$.

\noindent
Finally, note that, from the definition of $f_E(E^{\prime})$ in
(\ref{SemiAns}) and from (\ref{equivalence}):
\begin{equation}
\label{Qu}
Q^{(k)}(E,E^{\prime})=\frac{\hbar^N\nu_1...\nu_N\Gamma(N)}
{\sqrt{\pi}\Gamma(N-\frac{1}{2})(4\hbar\omega_RE
E^{\prime})^{N-1}}
\left[-(E^{\prime}-E)^2-\hbar^2\omega_R^2+2\hbar\omega_R(E^{\prime}+E)
\right]^{N-3/2}.
\end{equation}
This function is symmetric in $E$ and $E^{\prime}$ as in the quantum
mechanical case (see (\ref{averageFC})). In the limit in which this result
holds, we eventually find the explicit form of the ergodic rate equation by
inserting Eq.~(\ref{Qu}) into Eq.~(\ref{rateEnergy}). An explicit form of the
rate
equation can also be evaluated in the limit of $N\gg 1$ ions
\cite{Unpublished}.
\begin{figure}
\begin{center}
\epsfxsize=0.35\textwidth
\epsffile{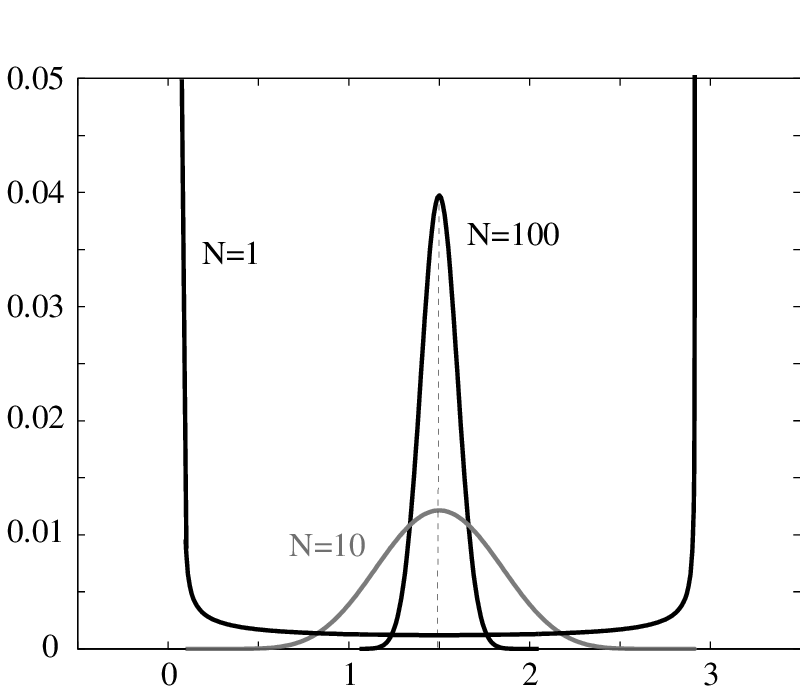} \caption{Plot of $g(E) Q^{(k)}(E,E^{\prime})$ as in
Eq.~(\ref{Qu}) as a function of $E$ for $N=1,10,100$. $E,E^{\prime}$ are
in arbitrary units. Here, $E=1$, $\hbar\omega_R=0.5$. The dashed line
indicates the location of the center of the distribution,
$E^{\prime}=E+\hbar\omega_R$.}
\end{center}
\end{figure}

\section{Conclusions}

\noindent We have studied Doppler cooling of a Coulomb crystal: Starting from
the full Master Equation, in the low saturation regime and by consecutive
steps of averaging we have derived a rate equation for the total energy of
the crystal, reducing dramatically the number of degrees of freedom and thus
simplifying considerably the complexity of the problem. The equation is
defined on a coarse-grained energy scale with grid $\Delta E$ such that
$\Delta E\ll\hbar\gamma$, where $\gamma$ is the linewidth of the electronic
transition resonant with laser light. Its derivation is based on the
quasi-continuum characteristic of the spectrum of motional energies $E$ on
$\Delta E$, and on the assumption that the coupling between states belonging
to different energy shells is a smooth function of $E$.

Starting from the general form of the equation, we have studied the
semiclassical limit and the Lamb-Dicke limit, and in both cases we have found
a Fokker-Planck equation describing cooling of a Coulomb crystal of $N$ ions.
The general solution agrees with the results of previous treatments, which
were developed in perturbation theory for these limiting cases
\cite{Stenholm,Lindberg,JavaSten,Javanainen}. As observed in
\cite{Javanainen}, the dynamics of cooling of a Coulomb crystal can be scaled
to the one of a single ion.

In the semiclassical limit we have derived the explicit form of the rate
equation for the mechanical energy, by calculating from the full quantum
mechanical expression the classical probability of scattering between
motional states at different energies. An explicit form of the equation can
also be derived in the quantum limit for the case of $N\gg 1$ ions. This
derivation will be presented in future work \cite{Unpublished}.

Our results, with marginal changes, can be applied to a Coulomb crystal in
three dimensions. In that case, the dimensionality enters into the number of
modes, which for a crystal of $N$ ions is $3N$, and into the spatial
distribution of the scattered photons, thus affecting the coefficients of the
energy rate equation. From the comparison between the cooling problem for a
one-dimensional crystal and for one ion in more dimensions, we expect the
final energy in 3D to differ from the one-dimensional result only by a
geometrical factor.

To conclude, we would like to remark on the generality of the treatment.
Other incoherent processes in physical systems can be studied in an analogous
way, provided that the rate determining the dynamics of interest can be
singled out, and that on the corresponding energy scale the spectrum of
energy levels is characterized by a quasi-continuum.

\section{Acknowledgements}

\noindent The authors are deeply grateful to H. Walther and P. Zoller, who have
motivated and stimulated this work in its various stages. Discussions with P. Zoller
have been highly appreciated. G.M. thanks J.I. Cirac, B.-G. Englert,
H. Hoffman, W. Schleich, and S. Stenholm for many stimulating
discussions and helpful comments. This work has been partly supported by the European
Commission (TMR networks ERB-FMRX-CT96-0077 and ERB-FMRX-CT96-0087) and by
the Austrian Science Fund FWF (SFB15).


\begin{appendix}

\section{Moments derivation}

\noindent {\bf One ion.} Let us consider a one-dimensional harmonic
oscillator of frequency $\nu$ and number state $|n\rangle$. Considering
$|\langle n|\exp(ikx)|k\rangle|^2$ as a distribution over the final states
$|k\rangle$ of energy $E_{\rm k}$, given the initial state $|n\rangle$ with
energy $E_{\rm n}$, the first and the second moments of the distribution are:
\begin{eqnarray}
\label{App:1a} &\langle {\mathcal E}_{\rm k}-{\mathcal E}_{\rm n}\rangle_{\rm
k}              & =\hbar\nu\sum_k (k-n)|\langle
n|\text{e}^{i\eta(a^{\dagger}+a)}|k\rangle|^2~,
\\
\label{App:1b} &\langle ({\mathcal E}_{\rm k}-{\mathcal E}_{\rm
n})^2\rangle_{\rm k}
                & =\hbar^2\nu^2\sum_k
(k-n)^2|\langle n|\text{e}^{i\eta(a^{\dagger}+a)}|k\rangle|^2.
\end{eqnarray}
Using the property $a^{\dagger}a|k\rangle=k|k\rangle$ and the
closure relation of the states $\{|n\rangle\}$, we can contract
the sum over $k$ in (\ref{App:1a}), (\ref{App:1b}), using:
\begin{equation}
\label{App:2} k^m|\langle
n|\text{e}^{i\eta(a^{\dagger}+a)}|k\rangle|^2= \langle
n|\text{e}^{-i\eta(a^{\dagger}+a)}\left(a^{\dagger}a\right)^m
\text{e}^{i\eta(a^{\dagger}+a)}|n\rangle.
\end{equation}
Expression (\ref{App:2}) can be further simplified by using the commutation
properties of the bosonic operators, and we obtain:
\begin{eqnarray}
&\langle {\mathcal E}_{\rm k}-{\mathcal E}_{\rm n}\rangle_{\rm k}
   &=\hbar\omega_R,\label{mf1}\\
&\langle ({\mathcal E}_{\rm k}-{\mathcal E}_{\rm n})^2\rangle_{\rm k}
   &=2\hbar\omega_R{\mathcal E}_{\rm n}+\hbar^2\omega_R^2,\label{mf2}
\end{eqnarray}
where we have used $\eta^2=\omega_R/\nu$.
\\

\noindent {\bf {\it N} ions}. Let us now take an $N$-ion chain. Using the
definitions of Section II, the first moment of the distribution is:
\begin{eqnarray}
\left( \langle {\mathcal E}_{\bf k}-{\mathcal E}_{\bf n}\rangle_{\bf k}
\right)_j &=&\sum_{\bf k}({\mathcal E}_{\bf k}-{\mathcal E}_{\bf n})
|\langle {\bf k}|\text{e}^{ikz_j}|{\bf n}\rangle|^2\nonumber\\
&=&\hbar\sum_{\alpha=1}^{N}\nu_{\alpha}\left(
k\sqrt{\frac{\hbar}{2m\nu_{\alpha}}}b_j^{\alpha}\right)^2
=\hbar\omega_R, \label{MN1}
\end{eqnarray}
where the subscript $j$ refers to the driven ion. Analogously, the second
moment has the form:
\begin{equation}
\label{Mom2}
\left( \langle \left({\mathcal E}_{\bf k}-{\mathcal E}_{\bf
n}\right)^2\rangle_{\bf k} \right)_j = \left(\frac{\hbar^2
k^2}{2m\nu}\right)^2\nu \sum_{\alpha=1}^{N}\nu_{\alpha}(2n_{\alpha}+1)
\left(b_j^{\alpha}\right)^2+\hbar^2\omega_R^2.
\end{equation}
Averaging (\ref{Mom2}) over the ions of the chain we find:
\begin{equation}
\langle \left({\mathcal E}_{\bf k}-{\mathcal E}_{\bf n}\right)^2\rangle_{\bf
k} = \frac{1}{N} \sum_j\left(\langle \left({\mathcal E}_{\bf k}-{\mathcal
E}_{\bf n}\right)^2\rangle_{\bf k}\right)_j
=2\frac{\hbar\omega_R}{N}{\mathcal E}_{\bf n}+\hbar^2\omega_R^2. \label{MN2}
\end{equation}

\section{Evaluation of the Semiclassical Trace}

\noindent Expression (\ref{ClassicTrace}) can be rewritten as:
\begin{equation}
F =\int_{{\cal F}(E)}\text{d}E_1\int_{{\cal F}(E^{\prime})}
\text{d}E_2\int\text{d}\bar{p}_1
\delta\left(E_2-E_1-\hbar\omega_R-\frac{\hbar
k}{m}\bar{p}_1\right) \int\text{d}\bar{q}_1...\text{d}\bar{q}_N
\frac{\text{d}\bar{p}_2...\text{d}\bar{p}_N}{h^N}
\delta\left(E_1-\sum_{\beta=1}^N\frac{\bar{p}_{\beta}^2}{2m}-
V(\bar{q}_1,..,\bar{q}_N)\right). \label{Effe}
\end{equation}
We define:
\begin{equation}
\label{Iintegral}
I(E,\bar{p}_1)=\int\text{d}\bar{q}_1...\text{d}\bar{q}_N
\frac{\text{d}\bar{p}_2...\text{d}\bar{p}_N}{h^N}
\delta(E-\sum_{\beta=1}^N\frac{\bar{p}_{\beta}^2}{2m}-
V(\bar{q}_1,..,\bar{q}_N)),
\end{equation}
and move to the coordinates ${q_{\beta}}$ where $V$ is a diagonal quadratic
form.
By introducing the set of rescaled variables
$Q_{\beta}=\sqrt{m\nu_{\beta}^2/2}q_{\beta}$,
$P_{\beta}=\sqrt{1/2m}p_{\beta}$, integral (\ref{Iintegral})
is the measure of the surface of a unitary hypersphere in $2N-1$
dimensions. Integrating, we obtain:
\begin{equation}
I(E,\bar{p}_1)
=\frac{2^{N}(E-\bar{p}_1^2/2m)^{N-3/2}}{\sqrt{2m}\nu_1...\nu_N}
\frac{\pi^{N-1/2}}{\Gamma(N-1/2)}.
\end{equation}
Substituting now this expression into (\ref{Effe}) and integrating
over $\bar{p}_1$ we get:
\begin{eqnarray}
F &=&\int_{{\cal F}(E)}\text{d}E_1\frac{1}{\Gamma(N-1/2)}
\int_{{\cal F}(E^{\prime})}\text{d}E_2\frac{m}{\hbar k}\frac{1}
{\sqrt{2\pi m}\hbar^N \nu_1...\nu_N} \left(E_1-\frac{(E_2-
\hbar\omega_R-E_1)^2}{4\hbar\omega_R}\right)^{N-3/2}\nonumber\\
&=&\int_{{\cal F}(E)}\text{d}E_1\frac{E_1^{N-1}} {\hbar^N
\nu_1...\nu_N\Gamma(N)}\int_{{\cal F}(E^{\prime})}
\text{d}E_2f_{E_1}(E_2),
\end{eqnarray}
where
\begin{equation}
\label{fee} f_E(E^{\prime})=\frac{\Gamma(N)}{\sqrt{\pi}\Gamma(N-1/2)}
\frac{1}{\sqrt{4\hbar\omega_RE}}
\left(1-\frac{(E^{\prime}-\hbar\omega_R-E)^2}{4\hbar\omega_RE}\right)^{N-3/2}.
\end{equation}
It can be easily verified that (\ref{fee}) satisfies the relations
(\ref{equivalence}) and (\ref{Prop}-\ref{Prop2}).
\end{appendix}


\begin{references}
\bibitem{LDRmany}
Note that the first order Lamb-Dicke expansion of each mode is consistent,
provided the condition $\omega_R/\nu_N\gg (\omega_R/\nu_1)^2$ holds (given
$\nu_N>\nu_{N-1}>...>\nu_1$).
\end{references}
\end{document}